\documentstyle[preprint,aps]{revtex}
\tightenlines
\def\ANP{Ann.\ Physics (N.Y.) }

\def\NPB{Nucl.\ Phys.\ }
\def\NPBP{Nucl.\ Phys.\ B (Proc.\ Suppl.) }
\def\PLB{Phys.\ Lett.\  } 
\def\PRL{Phys.\ Rev.\ Lett.\ }
\def\PRD{Phys.\ Rev.\  }

\def\MPL{Mod.\ Phys.\ Lett.\ }
\def\IJMP{Int.\ Jou.\ Mod.\ Phys.\ }

\def\JHEP{J.\ High Energy Phys.\ }

\def\e{\epsilon}
\def\f{\phi}
\def\g{\gamma}

\def\o{\omega}
\def\l{\lambda}
\def\la{\l^{2}}
\def\p{\pi}

\def\r{\rho}
\def\s{\sigma}
\def\t{\tau}

\def\ds{ds^2=}
\def\sg{\sqrt{-g}}

\def\sg{\sqrt{-g}}

\def\fo{\f_0}

\def\ord#1{O\left(#1\right)}

\def\r{\rho}
\def\s{\sigma}

\def\ds{ds^2=}
\def\t{\tau}

\def\be{\begin{equation}}
\def\ee{\end{equation}}
\def\bea{\begin{eqnarray}}
\def\eea{\end{eqnarray}}
\def\bc{\begin{displaymath}}
\def\ec{\end{displaymath}}
\def\lb{\label}
\def\ads{AdS}
\def\adsd{$\rm AdS_{2}$}

\def\adsdp{$\rm AdS_{2}^{+}$}

\def\adsdz{$\rm AdS_{2}^{0}$}
\def\be{\begin{equation}}
\def\ee{\end{equation}}
\def\ba{\begin{eqnarray}}
\def\ea{\end{eqnarray}}

\def\two{^{(2)}}
\def\alm{\gamma_m}
\def\talm{\tilde\gamma_m}
\def\bz{\bar z}
\def\p{\partial}
\def\bp{\bar\partial}
\def\pt{\partial_t}
\def\px{\partial_x}

\def\tchi{\tilde\chi}
\def\talpha{\tilde\alpha}
\def\summ{\sum_{m=-\infty}^{+\infty}}
\def\sumn{\sum_{n=-\infty}^{+\infty}}
\def\sumk{\sum_{k=0}^{+\infty}}
\def\bt#1#2{\beta_{#1,#2}}

\def\em{\epsilon_m}
\begin{document}
\preprint{\vbox{\noindent
INFNCA-TH0013  \null\hfill }}
%
\draft
\vskip 3truecm
\title{Open strings, 2D gravity and AdS/CFT correspondence}
\author{Mariano Cadoni$^{a}$\footnote{Email: cadoni@ca.infn.it},
Marco Cavagli\`a$^{b}$\footnote{Email: cavaglia@mercury.ubi.pt}}
\address{$^a$ Dipartimento di Fisica, Universit\`a di Cagliari,\\
Cittadella
Universitaria 09042, Monserrato, Italy and INFN, Sezione di Cagliari\\
$^b$ Departamento de F{\'\i}sica, Universidade da Beira Interior\\ 
R.\ Marqu\^es d'\'Avila e Bolama, 6200 Covilh\~a, Portugal.}
\vskip 3truecm
\maketitle
\begin{abstract}
We present a detailed discussion of the duality between dilaton gravity on
\adsd\ and open strings. The correspondence between the two theories is
established using their symmetries and field theoretical, thermodynamic, and
statistical arguments. We use the dual conformal field theory to describe
two-dimensional black holes. In particular, all the semiclassical features of
the black holes, including the entropy, have a natural interpretation in terms
of the dual microscopic conformal dynamics. The previous results are discussed
in the general framework of the Anti-de Sitter/Conformal Field Theory
dualities. 
\end{abstract}
%
\section{introduction}
One of the most striking features of the Anti-de Sitter/Conformal Field Theory
(\ads/CFT) correspondence \cite{Wm} is the possibility of relating physical 
theories that appear completely different at first sight. Although the meaning
of the \ads/CFT correspondence is yet to be fully clarified, we expect it will
help shed light on fundamental issues of contemporary theoretical physics, such
as the non-perturbative regime of Yang-Mills and string theories. 

Lower-dimensional models are often used in theoretical physics as simplified
models to investigating complex systems. This approach allows to formulate the 
problem under investigation in a mathematical simpler context, yet retaining
the crucial characteristics of the original model. Applying this strategy to
the AdS/CFT correspondence we are lead to investigate the lowest-dimensional,
$d=2$, member of the AdS$_{d}$/CFT$_{d-1}$ family. In this case the AdS/CFT
conjecture states that gravity on \adsd\ is dual to a one-dimensional conformal
field theory living on the timelike boundary of \adsd. Widely investigated in
the recent literature, the AdS$_{2}$/CFT$_{1}$ correspondence has, however,
revealed itself much more puzzling than its higher dimensional counterparts
\cite{g1,CM,CM3,CM1,g2,g3,g4,g5}. Specific features of two-dimensional gravity
and of the conjectured CFT living on the boundary of \adsd\ conspire indeed to
make the whole subject very difficult to analyze. Classical two-dimensional
(dilaton) gravity is a conformal theory itself. It can be formulated as a
nonlinear sigma-model \cite{cav1}, which, at the classical level, is endowed
with conformal symmetry. So we would naively expect gravity on \adsd\ to be
dual to a {\it two-dimensional} CFT. However, it has been shown that the
conformal symmetry associated with AdS$_2$ is infinite dimensional and is
generated by a  Virasoro algebra. Moreover, there is some evidence that it can
be realized in terms of boundary fields describing deformations of the boundary
of \adsd\  \cite{CM,CM3,CM1}. These results seem to indicate that the conformal
theory is actually a one-dimensional CFT, though the search for a viable
candidate has not been successful yet. (See Refs.\  \cite{CM,CM3,g3}.) 

The above features have a strong impact on the study of two-dimensional gravity
structures, i.e., black holes, by means of conformal field theory techniques.
Previous attempts to calculate the statistical entropy of  \adsd\ black holes
were only partially successful \cite{CM,CM3,CM1} (A mismatch of a factor $\sqrt
2$ between the thermodynamic and statistical entropy was found). Though the
free energy of \adsd\ black holes depends quadratically on the Hawking
temperature \cite{CM2}, a feature which is typical of two-dimensional CFTs, it
has been shown that \adsd\  black holes are completely characterized by the
charges associated with the asymptotic symmetries of \adsd\ \cite{CM,CM3}.
These charges are defined on the timelike boundary of \adsd, suggesting that
\adsd\ black holes admit a description in terms of  a one-dimensional conformal
field theory. 

In this paper we make a step forward in clarifying the meaning of the AdS/CFT
duality in two dimensions. Starting from the nonlinear sigma model description
of two-dimensional dilaton gravity \cite{cav1} we discuss in depth the duality
between two-dimensional dilaton gravity on \adsd\ and open strings. We show
that in the weak-coupling regime two-dimensional dilaton gravity on \adsd\ has
two different degeneration limits which correspond to Neumann and Dirichlet
boundary conditions for the open string, respectively. We put the modes of the
gravitational theory on the boundary in a one-to-one correspondence with the
string modes and explain the semiclassical properties of the \ads\ black hole
-- including the entropy -- in terms of the dual CFT microscopic  dynamics.
Some results of this paper have been anticipated in a previous letter
\cite{CC}. Here we extend and complete those results, in particular we present
a detailed and systematic discussion of the \adsd/CFT duality and clarify the
meaning of Dirichlet and Neumann boundary conditions. We also speculate on the
relevance of our results in the more general framework of  higher-dimensional
\ads/CFT dualities.

The structure of the paper is as follows. In Section II and Section III we
briefly review the main features of two-dimensional dilaton  gravity on \adsd\
and its formulation as a nonlinear sigma model, respectively. We also show that
in the weak-coupling regime the theory is described by an open bosonic string.
In Section IV we compare the symmetries of two-dimensional dilaton gravity on
\adsd\ to the symmetries of the string. In Section V we use the previous
results and further field theoretical arguments to show that gravity on \adsd\
is dual to the bosonic string. In Section VI we put in a one-to-one
correspondence the string modes with the asymptotic modes of \adsd\ gravity. 
In Section VII we use the \ads/CFT correspondence to explain the semiclassical
properties of the \adsd\ black hole in terms of the microscopic conformal
dynamics. Finally, in Section VIII we discuss our results.
\section {The two-dimensional dilaton gravity theory}
Our starting point is the two-dimensional dilaton gravity action
\be\lb{e1}
A={1\over2}\int d^2x\, \sg \, \left(\f R+V(\f)
\right).    
\ee
The scalar field $\phi$ is related to the usual definition of the dilaton
$\varphi$ by $\phi=\exp(-2\varphi)$. The two-dimensional model (\ref{e1}) has
been widely investigated in the literature \cite{dg}. Because of its simplicity
it has been used to address fundamental problems of quantum gravity and black
hole physics in a mathematically simplified context.

In this paper we restrict attention on dilaton gravity models that have \adsd\
as classical solution. The prototype of these models is the Jackiw-Teitelboim
theory \cite{JT}, JT for short, which is obtained setting $V(\f)=2\la\f$ in
Eq.\ (\ref{e1}). Although the JT theory may look rather uninteresting -- there
are no local physical degrees of freedom, the general solution describes a
spacetime of constant negative curvature -- a closer examination reveals a much
richer structure. In particular, the theory admits black hole solutions
\cite{CM2}. In the following we will briefly review the main features of 
\adsd\ black holes, referring the reader to the vast literature on the subject
for a more detailed discussion. (See, e.g., Ref.\ \cite{CM2} and references
therein.)

Owing to the extended Birkhoff theorem the general solution of the JT model in
the Schwarzschild gauge is
\be\lb{e2}
\ds-\left(\l^2r^2- {2m_{bh}\over \l \fo}\right)dt^2+\left(\l^2r^2-
{2 m_{bh}\over \l \fo}\right)^{-1}dr^2\,,
\quad \f=\fo \l r\,,\quad m_{bh}\ge 0\,.
\ee
The general theory (\ref{e1}) admits the existence of the gauge invariant,
local conserved quantity \cite{mass}
\be\lb{e3}
M=N(\phi)-g_{\mu\nu}\nabla^\mu\phi\nabla^\nu\phi\,,~~~~N(\phi)=\int^\phi
d\phi'V(\phi')\,.
\ee
On the classical orbit $M$ is constant and proportional to the ADM mass of the
system \cite{MTW}. For the JT black hole we have $M=2\fo \l  m_{bh}$. For
purely dimensional reasons two-dimensional dilaton gravity does not allow a
dimensionful analog of the four-dimensional Newton constant. However, $\f
^{-1}$ represents the (coordinate  dependent) coupling constant of the theory.
For the JT model, in particular, $\fo^{-1}$ plays the role of a dimensionless
Newton constant, $G_2$. The metric (\ref{e2}) represents different, locally
equivalent, parametrization of \adsd\ according to the value of $m_{bh}$. The
presence of the scalar field $\phi$ makes these parametrizations globally
inequivalent \cite{CM2}. In this paper, following the notations of Ref.\
\cite{CM2}, solutions with  $m_{bh}>0$ and $m_{bh}=0$ will be denoted by
\adsdp\ and \adsdz, respectively. \adsdp\ can be interpreted as a black hole of
mass $m_{bh}$ with a singularity at $r=0$, a  timelike boundary at $r=\infty$,
and an event horizon at $r=(2 m/\l^{3}\fo)^{1/2}$. \adsdz\ can be considered as
the ground state, zero mass solution. In this case the singularity at $r=0$ is
lightlike. Let us stress that the global topology of both \adsdz\ and \adsdp\
is different from the topology of the full AdS$_{2}$ geometry (the maximally
extended spacetime). The latter is a geodesically complete spacetime with
cylindrical topology and two timelike boundaries. Because of the singularity at
$r=0$ both \adsdp\ and \adsdz\ are singular spacetimes with a single timelike
boundary at $r=\infty$.

 Since \adsdp\ and \adsdz\ are locally equivalent, a coordinate transformation exists
that maps the solution (\ref{e2}) with $m_{bh}>0$ into the solution with
$m_{bh}=0$ \cite{CM2}. Later on this paper we will make use of this coordinate
transformation. In the conformal gauge the \adsdz\ metric is
\be\lb{e4}
ds^{2}={1\over \l^2x^2} (-dt^2 +dx^2)\,.
\ee
The metric of the \adsdp\ black hole is
\be\label{e5}
ds^2={a^{2}\over\sinh^2(a\l\s)}(-d\tau ^2+d\s^2)\,,
\ee
where $a=(2 m_{bh}/\fo \l)^{1/2}$. The two metrics above are related by the
change of coordinates
\be\label{e6}
t={1\over a \l}e^{a\l\t}\cosh(a\l\s)\,,\qquad
x= {1\over a \l} e^{a\l \t}\sinh(a\l\s)\,.
\ee
In the following, we will also use the light-cone coordinates
\be\label{e6a}
u={1\over 2}(t+x),\qquad v={1\over 2}(-t+x)\,.
\ee
In this coordinate frame the \adsdz\ solution is
\be\lb{e6b}
ds^{2}={4\over \l^2(u+v)^2} dudv,\qquad \phi=-\fo [\l(u+v)]^{-1}.
\ee
The black hole solution (\ref{e2}) can be interpreted as a thermodynamic system
and the usual thermodynamic parameters can be associated to it. The black hole
mass depends quadratically on both the Hawking temperature $T$ and the entropy
$S_{bh}$ \cite{CM2},
\be\label{e7}
 m_{bh}¥={2\pi^{2}\fo\over \l} \, T^{2}\,,\qquad
 S_{bh}=4 \pi \sqrt{m \fo\over 2 \l}\,.
\ee
A fundamental question concerns the statistical interpretation of the 
thermodynamic quantities in Eq.\ (\ref{e7}). In particular, one would be
able to identify the microscopic degrees of freedom whose dynamics produces
the huge degeneracy which is contained in Eq.\ (\ref{e7}).

At the semiclassical level black holes are unstable because of the Hawking
effect. In the two-dimensional context the Hawking evaporation process has a
simple and nice explanation. From the coordinate transformation (\ref{e6}) we
find that the relation between \adsdp\ and \adsdz\ is formally equivalent  to
the relation between Rindler and Minkowski spacetimes. By quantizing a scalar
field in the fixed backgrounds defined by \adsdp\ and \adsdz\ one finds that
the positive frequency modes of the quantum field with respect to Killing
vector $\partial_{t}$ are not positive frequency modes with respect to Killing
vector $\partial_{\t}$. Hence, the vacuum state which is seen by an observer in
the $(\t,\s)$ reference frame appears filled with thermal radiation to an
observer in the $(t,x)$ frame. The flux corresponds to a Planck spectrum with
temperature given by Eq.\ (\ref{e7}) \cite{CM2}. The  relation between mass and
temperature in Eq.\ (\ref{e7}) is nothing else but the two-dimensional
Stefan-Boltzmann law.

The previous features strongly suggest the existence of an underlying
two-dimensional field theory whose microscopic dynamics is responsible for the
thermodynamic behavior of the black hole.
\section{The sigma model approach to two-dimensional dilaton gravity}
The dilaton gravity action (\ref{e1}) can be cast in a nonlinear conformal
sigma model form \cite{cav1}. The two-dimensional Ricci scalar $R^{(2)}(g)$ can
be locally written as
\be\label{e8}
R^{(2)}(g)=2\,\nabla_\mu A^\mu\,,~~~~A^\mu=
{\nabla^\mu\nabla^\nu\chi\nabla_\nu\chi-
\nabla_\nu\nabla^\nu\chi\nabla^\mu\chi\over
\nabla_\rho\chi\nabla^\rho\chi}\,,
\ee
where $\chi$ is an auxiliary scalar function. Equation (\ref{e8}) can be
checked using conformal coordinates and general covariance arguments.
Differentiating Eq.\ (\ref{e3}), setting $\chi=\phi$ in Eq.\ (\ref{e8}), and
integrating per parts, the action (\ref{e1}) can be written as a functional of
$M$ and $\phi$
\be\label{e9}
A={1\over 2}\int_\Sigma d^2x\, \sg \, {\nabla_\mu\phi\nabla^\mu M
\over N(\phi)-M}\,.
\ee
Clearly, the action (\ref{e9}) describes a two-dimensional nonlinear sigma
model. In the canonical form, using the metric parametrization
\be\label{e10}
g_{\mu\nu}=\rho\left(\matrix{\alpha^2-\beta^2&\beta\cr
\beta&-1\cr}\right)\,,
\ee
the super-Hamiltonian and super-momentum are
\ba\label{e11}
{\cal H}_0&=&2[N(\phi)-M]\pi_{\phi}\pi_M+{1\over 2
[N(\phi)-M]}\phi'M'\,,\\
{\cal H}_1&=&-\phi'\pi_{\phi}-M'\pi_M\,.
\ea
The canonical action must be complemented by a boundary term at the spatial
boundaries to make the action finite and differentiable. According to Ref.\
\cite{RT} the boundary term coincides with the conserved charge, i.e., the
mass, of the black hole. [See below, Eq.\ (\ref{f6})]. The canonical chart
$(\phi,\pi_{\phi},M,\pi_M)$ is related to the canonical chart
$(\phi,\Pi_{\phi},\rho,\Pi_\rho)$ by the map
\ba\label{e12}
M&=&N(\phi)-\displaystyle{4\rho^2\Pi_\rho^2-\phi'^2\over\rho}\,,\nonumber\\
\pi_M&=&\displaystyle{\rho^2\Pi_\rho\over 
4\rho^2\Pi_\rho^2-\phi'^2}\,,\\
\pi_{\phi}&=&\Pi_\phi-\displaystyle{\rho^2\Pi_\rho\over
4\rho^2\Pi_\rho^2-\phi'^2}\left[V(\phi)+2\Pi_\rho\left({\phi'\over
\rho\Pi_\rho}\right)'\right]\,.\nonumber
\ea
Equation (\ref{e12}) proves the equivalence of Eq.\ (\ref{e1}) and Eq.\
(\ref{e9}) at canonical level.

Let us consider the JT model. In this case it is convenient to define the
``coupling constant" field
\be\label{e13}
\psi=-{1\over 2\lambda^2\phi}\,.
\ee
In terms of $M$ and $\psi$ the JT action is
\be\label{e14}
A=\int d^2x\, \sg \, \partial_\mu
M\partial^\mu\psi\cdot{1\over 1-4\lambda^2\psi^2\,M}\,.
\ee
The boundary of the spacetime is now located at $\psi=0$. The action (\ref{e14})
can be expanded around $\psi=0$,
\be\lb{e15}
A=\int d^2x\, \sg \, \partial_\mu M\partial^\mu\psi
\left[1+\sum_{k=1}^{+\infty}(2\lambda)^{2k}M^k\psi^{2k}\right]\,.
\ee
Equation (\ref{e15}) is both a weak-coupling expansion in terms of the
coordinate-dependent gravitational coupling of the model and an expansion near
the boundary of \adsd. This fact suggests that the gravitational theory can be
represented perturbatively as an expansion around the boundary. The first term
(zero order) of the expansion coincides with the action for a bosonic string
living in a two-dimensional flat target spacetime and describes the (off-shell)
weak-coupled gravitational theory. It can be cast in the usual form (see Ref.\
\cite{pol} for notations)
\ba\lb{e16}
A_0&=&{1\over 2\pi\alpha'}\int d^2z\partial X^\mu\bar\partial 
X_\mu\nonumber\\
&=&{1\over 2\pi\alpha'}\int d^2z(-\partial X^0\bar\partial 
X^{0}+\partial X^1\bar\partial X^1)\\
&=&{1\over 2\pi\alpha'}\int d^2z(\partial X^2\bar\partial 
X^{2}+\partial X^1\bar\partial X^{1})\,,\nonumber
\ea
by defining the new fields 
\be\lb{e17}
\begin{array}{lllll}
\sqrt{\pi\alpha'}M&=&\displaystyle{1\over 2}(X^1+iX^2)&=&\displaystyle
{1\over 2}(X^1+X^0)\,,\\\\
\sqrt{\pi\alpha'}\psi&=&\displaystyle{1\over 2}(X^1-iX^2)&=&\displaystyle
{1\over2}(X^1-X^0)\,,
\end{array}
\ee
where $\sqrt{\alpha'}$ is the string length, and
\be\lb{e18}
z\equiv u={1\over 2}(\sigma^1+i\sigma^2)\,,\qquad
\bar z\equiv v={1\over 2}(\sigma^1-i\sigma^2)\,.
\ee
Higher orders in the expansion (\ref{e15}) can be interpreted as interaction
terms for the bosonic string (\ref{e16}). They describe perturbative
(off-shell) effects induced by the gravitational bulk on the boundary.
Classically, the two-dimensional JT model is a topological theory with no
propagating physical degrees of freedom. Owing to the Birkhoff theorem
\cite{cav1} the physics on the gauge shell, i.e., in the fundamental state, is
completely determined by the spacetime boundary where the conserved charge is
defined, whereas the bulk is pure gauge. In the perturbative sigma model
approach the first term of the expansion (\ref{e15}) can be interpreted as
describing both  the (off-shell) gravitational theory on the boundary and the
weak-coupling regime of the theory. Therefore, we expect that the free bosonic
string (\ref{e16}) -- with properly fixed boundary conditions -- describes the
semiclassical properties of the theory. Higher orders in the coupling constant
perturbative expansion (\ref{e15}) describe the corrections to the off-shell
dynamics on the boundary and lead, in the quantum theory, to gravitational
corrections to the classical geometry. (See e.g.\ Ref.\ \cite{cav2} where
quantum corrections to the ADM mass of the Schwarzschild black hole have been
calculated at the second order in the curvature expansion.) 

The boundary conditions to be imposed on the free bosonic string (\ref{e16})
are essential in determining the physical content of the AdS/CFT
correspondence. First of all, we note that AdS$_{2}$ has a timelike boundary at
$x=0$, so the CFT (\ref{e16}) must necessarily describe open strings. Since
open strings propagating in a two-dimensional target spacetime do not have
transverse excitations, we can impose either Dirichlet boundary conditions
[$\partial_{a}X^{\mu}(x=0)=0$] or Neumann boundary conditions
[$n^{a}\partial_{a} X^{\mu}(x=0)=0$, where $n^{a}$ is the normal to the
boundary]. Expanding the fields on the boundary [see below Eq.\ (\ref{f7})], 
in the former case we have
\be\lb{e19}
X^{\mu}(x=0)=\sqrt{\pi\alpha'}M_0(t)=\hbox{constant}\,,
\ee
where $M_0(t)$ is the (constant) zero-mode of the mass field $M$ on the
boundary. Dirichlet boundary conditions break translation invariance. Moreover,
they hold fixed the endpoint of the string on the boundary and do not allow any
dynamical degree of freedom on the boundary itself. Hence, Dirichlet boundary
conditions realize a AdS$_{2}$/CFT$_{2}$ correspondence. The
one-dimensional boundary can be interpreted as a $D$-brane ($0$-brane).
Possibly, a non-trivial dynamics on the brane can be generated by the
introduction of Chan-Paton factors. (See, e.g., Ref.\ \cite{pol}.)

Neumann boundary conditions do not break translation invariance. They allow for
excitations on the boundary, 
\be\lb{e20}
X^{\mu}(x=0)=\sqrt{\pi\alpha'}M_0(t)\,.
\ee
Since Neumann boundary conditions allow dynamical degrees of freedom on the
boundary, they seem to realize a AdS$_{2}$/CFT$_{1}$ correspondence, where
CFT$_{1}$ is a genuine one-dimensional CFT generated by the charges living on
the boundary \cite{CM,CM3}.

In addition to the timelike boundary at $x=0$,  AdS$_{2}^{0}$ has an inner null
boundary. However, the presence of the latter does not influence the dynamics
of the open string. In the conformal coordinate frame $(t,x)$ the metric of
\adsdz\ is given by Eq.\ (\ref{e4}) and the presence of the dilaton requires
\be\lb{e22}
-\infty<t<\infty\,,\qquad 0\le x<\infty\,.
\ee
In this coordinate frame the inner null boundary is located at $x=\infty$. Eq.\
(\ref{e4}) implies that \adsdz\ is conformal to the Minkowski spacetime. Hence,
because of conformal invariance open strings on \adsdz\ are equivalent to open
strings on the region of the $(t,x)$ Minkowski spacetime defined by Eq.\
(\ref{e22}). In the next section we will discuss how the symmetries of the
bosonic string (\ref{e16}) reflect in the asymptotic symmetries of the
two-dimensional gravitational theory (\ref{e1}).
\section{Symmetries of two-dimensional gravity and symmetries of the string}
\adsd\ is a maximally symmetric space, so the JT theory admits three Killing
vectors that generate the $SO(1,2)\sim SL(2,R)$ group of isometries.  In the JT
theory the presence of the dilaton actually breaks the $SL(2,R)$ symmetry
\cite{CM1}. However, this is irrelevant for the present discussion. Indeed, in
the weak-coupling regime, $\psi\to 0$, one naturally expects the $SL(2,R)$
symmetry to be enlarged to the full asymptotic symmetry group of \adsd. Since
Eq.\ (\ref{e15}) is a near-boundary expansion, the only relevant symmetries are
the symmetries that leave the \adsd\ metric asymptotically invariant.

The symmetries of the bosonic string (\ref{e16}) are related to the asymptotic
symmetry group of \adsd. The latter has been studied in detail in Refs.\
\cite{CM,CM3}. The asymptotic symmetries of \adsd\ can be found by imposing
suitable boundary conditions for the metric at $r\to \infty$. These boundary
conditions express the intuitive notion of ``asymptotically anti-de Sitter''
and allow the charges associated with the symmetry to be properly defined.

In the Schwarzschild gauge the boundary conditions to be imposed on the metric
$ds^2=g_{\mu\nu}dx^\mu dx^\nu$ and on the scalar field $\phi$ are 
\cite{CM,CM3}
\be\lb{f1a}
\begin{array}{lcl}
g_{tt}&=&\displaystyle -\l^2r^2+\g_{tt}+\ord{1\over r}\,,\\
g_{tr}&=&\displaystyle {\g_{tr}\over\l^3 r^{3}}+
\ord{1\over x^{5}}\,, \nonumber\\
g_{rr}&=&\displaystyle {1\over\l^2r^2}+{\g_{tt}\over\l^4 r^4}+\ord{1\over r^{6}}\,,
\end{array}
\ee
and 
\be\lb{f1b}
\f=\fo \left[\l r+ \r \l r+ {\g_{\f\f}\over \l r}+ \ord{1\over 
r^{2}}\right]\,,
\ee
respectively. In the previous equations the $\g$'s and $\rho$ are arbitrary
functions of $t$ and can be thought as characterizing the deformations of the 
boundary of \adsd\ and of the dilaton field. In the conformal gauge the
boundary of \adsd\ is located at $u=-v$ and the above conditions (at the order
$k$) read
\be\lb{f2}
\begin{array}{lcl}
g_{uu}&=&U_0(u-v)+\ldots +U_k(u-v)(u+v)^{k}+O[(u+v)^{k+1}]\\\\
g_{uv}&=&\displaystyle {2\over \lambda^2(u+v)^2}+Y_0(u-v)+\ldots +Y_{k}(u-v)(u+v)^{k}+
O[(u+v)^{k+1}]\\\\
g_{vv}&=&V_0(u-v)+\ldots +V_k(u-v)(u+v)^{k}+O[(u+v)^{k+1}]\\\\
\phi&=&\displaystyle -\phi_0\left[{\omega_{-1}\over \lambda
(u+v)}+\omega_{1}\lambda(u+v)+\ldots+ 
\o_{k}\lambda^{k}(u+v)^{k}+O[(u+v)^{k+1}]\right]
\end{array}
\ee
where the coefficients $\Theta_{k}=(U_{k},V_{k},Y_{k},\omega_{k})$ are
arbitrary functions of $u-v$. By definition the leading terms in Eqs.\
(\ref{f2}) are invariant under the transformations generated by the asymptotic
symmetry group. The functions $\Theta_{k}$ change according  to a
representation of the asymptotic symmetry group. Solving the Killing equations
for the metric (\ref{f2}), we find that the asymptotic symmetry group is
generated by the Killing vectors
\be\lb{f3}
\chi^{AdS}=\chi^u(u,v)\p_u+\chi^v(u,v)\p_v\,,
\ee
where
\be\lb{f3ab}
\chi^u={1\over 2}\left[ \epsilon+\epsilon'(u+v)+
{1\over 2}\epsilon''(u+v)^2 \right]+ \alpha^u\,,\quad
\chi^v= {1\over 2}\left[-\epsilon+\epsilon'(u+v)- 
{1\over 2}\epsilon''(u+v)^2 \right]+\alpha^v\,.
\ee
Here, $\epsilon$ is an arbitrary function of $u-v$, $'$ denotes differentiation
with respect to $u-v$, and $\alpha^{u,v}= \sum_{k=3}^{+\infty}
\alpha^{u,v}_{k}(u-v) (u+v)^{k}$. The functions $\alpha^{u,v}$ represent ``pure
gauge'' diffeomorphisms of the two-dimensional gravitational theory that fall
off rapidly on the boundary. Expanding the function $\epsilon(u-v)$ in power
series, the Killing vectors (\ref{f3}) are recognized to define a conformal
group which is generated by the Virasoro algebra
\be\lb{f4}
[L^{AdS}_m,L^{AdS}_n]=(m-n)L^{AdS}_{m+n}\,.
\ee
The boundary fields $\Theta_{k}$ span a representation of the conformal group.
Their transformation law is
\be\lb{f5}
\delta_\e \Theta_{k}= \epsilon \Theta_{k}'+(h+k)\epsilon'\Theta_{k}+\ldots\,,
\ee
where dots denote terms that depend on higher derivatives of $\e$ and on pure
gauge diffeomorphisms, and $h=2$ for $U_k,V_k,Y_k$ and $h=0$ for $\omega_k$,
respectively. Note that the pure gauge transformations affect the boundary
fields but leave invariant the charge associated with the falloff conditions,
\be\lb{f6}
J(\e)=\epsilon \l\fo\left[2\o_{1}+{1\over
8}(U_{0}+V_{0}+2Y_{0})\right]=\epsilon{M_0(t)\over 2\lambda\phi_0}\,,
\ee
where $J(\e)$ has been calculated for $\o_{-1}=1$. Both the mass functional $M$
and the coupling constant field $\psi$ can be expanded in power series around
the boundary
\be\lb{f7}
M=\sumk M_{k}(u-v) (u+v)^{k}\,,\qquad
\psi=\sum_{k=1}^{+\infty} \psi_{k}(u-v) (u+v)^{k}\,.
\ee
Using Eqs.\ (\ref{e3}) and (\ref{e13}) both $M_{k}$ and $\psi_k$ can be 
expressed in terms of the boundary fields. They transform according to Eq.\
(\ref{f5}) with $h=0$. The two-dimensional dilaton gravity action or,
alternatively, the sigma model action can be expanded around the boundary as
well. Expanding in power series the Lagrangian density, ${\cal
L}=\sum_{k=0}^{+\infty} {\cal L}_k (u-v)(u+v)^k$, we find that ${\cal L}_k$
transform according to Eq.\ (\ref{f5}) with $h=2$, as is expected for a
two-dimensional conformal field theory.  

The sigma model action (\ref{e9}) is classically invariant under the conformal
transformations of the two-dimensional world-sheet. This invariance is not
manifest in its two-dimensional gravitational counterpart (\ref{e1}). Conformal
invariance of two-dimensional dilaton gravity is more subtle and can be
understood in terms of the Weyl-rescaling invariance of the target space 
coordinates $\psi$ and $M$ \cite{Cadoni}. At the leading order in the
$\psi\to 0$ expansion the conformal symmetry of the sigma model is the usual
two-dimensional conformal symmetry group of the free bosonic string (\ref{e16})
which is generated by the Killing vectors
\be\lb{f8}
\chi^{CFT}=\chi(z)\,\p+\tchi(\bz)\,\bp\,.
\ee
The transformation law of a generic $CFT_{2}$ field $X(z,\bar z)$ of weights
$(h,\tilde h)$ is
\be\lb{f9}
\delta_{\chi,\tchi}X= (\chi\p+h\p\chi)X+
(\tchi\bp+\tilde h\bp\tchi)X.
\ee   
Expanding $\chi(z)$ and $\tchi(\bz)$ as
\be\lb{f10}
\chi(z)=\summ\alm z^{-m+1}\,,\qquad
\tchi(\bar z)=\summ\talm \bar z^{-m+1}\,,
\ee
we have
\be\lb{f11}
\chi^{CFT}=\summ\left(\alm L^{CFT}_{m}+\talm\tilde L^{CFT}_{m}\right)\,,
\ee
where
\be\label{f12}
L^{CFT}_{m}=z^{-m+1}\p\,,\qquad \tilde L^{CFT}_{m}=\bz^{-m+1}\bp\,,
\ee
each satisfy the Virasoro algebra (\ref{f4}). Finally, the stress-energy tensor
is
\be\label{f13}
T_{zz}=-{1\over 2\pi \alpha'}\p X^\mu\p X_\mu=-2\p M\p\Psi=
{1\over 2\pi}\summ L^{CFT}_{m} z^{-2-m}\,.
\ee

In the next section we will see that the asymptotic symmetry group of \adsd\
(with fixed pure gauge diffeomorphisms) coincides with the conformal symmetry
group of the free bosonic string with properly chosen boundary conditions. 
\section{Duality of two-dimensional gravity on AdS$_2$ and open strings}
The duality between gravity on AdS$_{2}$ and the open string can be realized by
putting in a one-to-one correspondence the symmetries of the string and the
asymptotic symmetries of AdS$_{2}$. The physical content of the AdS$_2$/CFT
correspondence varies according to the boundary conditions that are chosen for
the bosonic string (\ref{e16}). We will consider first Dirichlet  and then
Neumann boundary conditions.
\subsection{Dirichlet boundary conditions}
Equations (\ref{f3ab}) suggest that the $u$ and $v$ components of the Killing
vectors of the asymptotic symmetry group, $\chi^u$ and $\chi^v$, are not
independent. Let us neglect initially the pure gauge diffeomorphisms in Eqs.\
(\ref{f3ab}) and define two auxiliary functions $\chi(u)$ and $\tchi(v)$
that satisfy the relation (notations will be clear soon) 
\be\lb{g1}
\chi((t+x)/2)|_{x=0}=-\tchi((-t+x)/2)|_{x=0}={1\over 2}\epsilon(t)\,.
\ee
At the second order in the expansion the $u$ and $v$ components of the Killing
vectors (\ref{f3}) can be recognized to be the first three terms of the
expansion around $x=0$ of the functions $\chi(u)$ and $\tchi(v)$
\be\lb{g2}
\chi={1\over 2}\sumk {1\over k!}{d^{k}\epsilon\over 
d(u-v)^{k}}(u+v)^{k}\,,\qquad
\tchi=-{1\over 2}\sumk (-1)^{k}{1\over k!}{d^{k}
\epsilon\over d(u-v)^{k}}(u+v)^{k}\,.
\ee
Assuming that the equivalence is valid at any order and taking into  account
Eq.\ (\ref{e18}), we find that the CFT$_{2}$ Killing vectors (\ref{f8})
coincide with the \adsd\ Killing vectors (\ref{f3}) where the gauge
diffeomorphisms have been fixed as
\be\lb{g3}
\alpha^{u}_{k}=(-1)^{k+1}\alpha^{v}_{k}={1\over 2k!}{d^{k}\epsilon\over
d(u-v)^{k}}.
\ee
Hence, the asymptotic symmetry group of AdS$_{2}$ coincides with the symmetry
group of the bosonic open string (\ref{e16}). In order to obtain the second one
from the first one we need to fix the gauge diffeomorphisms of the
gravitational theory. This gives a non trivial relation between the
diffeomorphisms of the two-dimensional dilaton gravity theory and the
diffeomorphisms of the conformal field theory.

The previous equations follow from the Dirichlet condition for the function
${\cal X}(u,v)\equiv\chi(u)+\tchi(v)$, i.e.,
\be\lb{g4}
{\cal X}(u,v)\equiv\chi(u)+\tchi(v)=0\,,
\ee
on the boundary $u+v=0$. This equation implies that the conformal symmetry is
generated by a single copy of the Virasoro algebra and the weights $(h,\tilde
h)$ appear in the transformation law (\ref{f9}) only in the combination
$w=h+\tilde h$. 

The correspondence between the conformal group of the bosonic string and the
asymptotic \adsd\ group can be observed directly on the \adsd\ fields
$\Theta_{k}$, $M_{k}$, $\psi_{k}$ and ${\cal L}_{k}$. Each of these fields can
be interpreted as the coefficient of the expansion of the corresponding CFT$_2$
field around the boundary with given weight $w=h+\tilde h$ and pole of order
$p$. We have the table:
\be
\vbox{
\offinterlineskip
\halign{
\strut
\vrule height 12pt#&\hfil\quad#\quad\hfil&\vrule#&\hfil\quad#\quad\hfil&\vrule#\
&\hfil\quad#\quad\hfil&#\vrule&\hfil\quad#\quad\hfil&\vrule#
&\hfil\quad#\quad&\vrule#\cr
\noalign{\hrule}
&\adsd\ Field&&$h$&&$\tilde h$&&$w$&&$p$&\cr
\noalign{\hrule}
&$U_{k}$&&$2$&&$0$&&$2$&&$0$&\cr
&$V_{k}$&&$0$&&$2$&&$2$&&$0$&\cr
&$Y_{k}$&&$1$&&$1$&&$2$&&$-2$&\cr
&$\omega_{k}$&&$0$&&$0$&&$0$&&$-1$&\cr
&$M_{k}$&&$0$&&$0$&&$0$&&$0$&\cr
&$\psi_{k}$&&$0$&&$0$&&$0$&&$1$&\cr
&${\cal L}_{k}$&&$1$&&$1$&&$2$&&$0$&\cr}
\hrule}
\ee
The table above represents the AdS$_{2}$/CFT$_{2}$ correspondence in terms of
fields: the \adsd\ fields are interpreted as objects of the two-dimensional
CFT. Using Eq.\ (\ref{f9}) together with Eqs.\ (\ref{g2}) we recover the
transformation (\ref{f5}) of the \adsd\ fields. 

The AdS$_{2}$/CFT$_{2}$ correspondence allows to determine the Virasoro
generators of the asymptotic symmetry group of AdS$_{2}$ from CFT$_2$. Let us
consider the Virasoro generators (\ref{f12}). Using the $(t,x)$ coordinates we
have
\be\lb{g5}
L^{CFT}_{m}=2^{m-1}\sumk t^{-m+1-k}x^k \bt{m-2}{k}(\pt+\px)\,,
\ee
where 
\be\lb{g6}
\bt{m-2}{k}=\left(\matrix{1-m\cr k\cr}\right)\,.
\ee
A similar expression holds for $\tilde L^{CFT}_{m}$. We have seen that the
asymptotic symmetry group of \adsd\ can be obtained from the conformal group of
the string by imposing Eq.\ (\ref{g4}). Applying the condition (\ref{g4}) to
Eq.\ (\ref{f11}) we find $\alm=(-1)^m\talm$. From Eq.\ (\ref{g5}) the Killing
vectors for the Dirichlet boundary conditions read
\be\lb{g7}
\chi^{CFT}=\summ\alm L'_m\,,
\ee
where
\be\lb{g8}
L'_m=2^m\sumk\left[\bt{m-2}{2k}t^{-m+1-2k}x^{2k}\pt+\bt{m-2}{2k+1}t^{-m-2k}
x^{2k+1}\px\right]\,.
\ee
The Killing vectors (\ref{g7}) coincide with the Killing vectors of the
asymptotic symmetry group of \adsd\ (\ref{f3}) with fixed pure gauge
diffeomorphisms. [See Eq.\ (\ref{g3}).] Indeed, let us consider Eq.\
(\ref{f3ab}) with fixed gauge diffeomorphisms and expand $\epsilon(t)$ in power
series 
\be\lb{g8b}
\epsilon(t)=\summ 2^m\epsilon_m t^{-m+1}\,.
\ee
The Killing vectors (\ref{f3}) are cast in the form
\be\lb{g9}
\chi^{AdS}=\summ\em L^{AdS}_m\,,
\ee
where
\ba\lb{g10}
L^{AdS}_m&=&2^m\left\{\left[t^{-m+1}+{1\over
2}(-m+1)(-m)t^{-m-1}x^2+\dots\right]\pt
+\left[(-m+1)t^{-m}x+\dots\right]\px\right\}\nonumber\\
&=&L'_m\,.
\ea
Setting $\em=\alm$ the Killing vectors of the \adsd\ asymptotic symmetry group
and the Killing vectors of the string with Dirichlet boundary conditions
coincide. Therefore, they generate both the full symmetry group of the bosonic
open string and the gravitational asymptotic symmetry group of \adsd\ with
fixed pure gauge diffeomorphisms. The Virasoro generators of the \adsd\
asymptotic symmetry group are
\be\lb{g11}
L^{AdS}_{m}=2^{m-1}\left\{\left[(t+x)^{-m+1}+(t-x)^{-m+1}\right]\pt
+\left[(t+x)^{-m+1}-(t-x)^{-m+1}\right]\px\right\}\,.
\ee
The Virasoro generators (\ref{g11}) are simply obtained from the Virasoro
generators of the string by taking the linear combination
$L^{AdS}_m=L^{CFT}_{m}+(-1)^m\tilde L^{CFT}_{m}$ and changing coordinates to
$(t,x)$. This relation implies that the symmetries of the open string are
generated by a single copy of the Virasoro algebra. On the boundary the
Virasoro generators are $L^{AdS}_{m}|_{x=0}=2^{m}t^{-m+1}\pt$.

It should be noted that both the Killing vectors and the Virasoro generators
are now defined outside the spacetime boundary. By fixing the gauge 
diffeomorphisms we select a subgroup of the full two-dimensional diffeomorphism
group of the gravitational theory. This subgroup is recognized to be the
conformal group of the string with Dirichlet boundary conditions. The latter
can also be defined as a subgroup of the diffeomorphisms that leave the
two-dimensional metric asymptotically invariant (\adsd/CFT$_{2}$ duality).

Finally, let us conclude this section with a few equations that will be useful
in the following. The subalgebra $SL(2,R)$ of the Virasoro algebra is generated
by
\be\lb{g12}
L^{AdS}_{0}=t\pt+x\px\,,\qquad
L^{AdS}_{1}=2\pt\,,\qquad
L^{AdS}_{-1}={1\over 2}(t^2+x^2)\pt+xt\px\,.
\ee
The Virasoro generator $L^{AdS}_{0}$ does not generate translations in $t$ but
dilatations. Changing coordinates to $(\tau,\sigma)$ [see Eq.\ (\ref{e6})] the
Virasoro generators become
\be\lb{g13}
L^{AdS}_{m}=(2a\l)^{m}e^{-ma\l\t}{1\over a\l}
\left[\cosh(ma\l\s)\p_\tau-\sinh(ma\l\s)\p_\sigma\right]\,.
\ee
In this reference frame $L^{AdS}_{0}$ generates translations in the new time
coordinate $\tau$. 
\subsection{Neumann boundary conditions}
Let us now discuss the AdS$_2$/CFT duality when Neumann boundary conditions are
enforced. Imposing the Neumann boundary conditions on the function ${\cal
X}(u,v)$ [see Eq. (\ref{g4})] we have, on the boundary $u+v=0$,  
\be\lb{g14}
\p_{u}\chi(u)+\p_{v}\tchi(v)=0\,.
\ee
Equation (\ref{g14}) is solved by the condition
\be\lb{g15}
\chi((t+x)/2)|_{x=0}=\tchi((-t+x)/2)|_{x=0}={1\over 2}\epsilon(t)\,.
\ee
Using Eq.\ (\ref{g15}) we try and put in a one-to-one correspondence the
symmetry group of the open string with Neumann boundary conditions with the
asymptotic symmetry group of \adsd. Imposing the condition (\ref{g15}) on the
CFT Killing vectors (\ref{f11}) we find $\alm=-(-1)^m\talm$. The Virasoro
generators of the asymptotic \adsd\ group with Neumann boundary conditions are
\be\lb{g16}
L^{AdS}_{m}=2^{m-1}\left\{\left[(t+x)^{-m+1}-(t-x)^{-m+1}\right]\pt
+\left[(t+x)^{-m+1}+(t-x)^{-m+1}\right]\px\right\}\,.
\ee
From the previous equation it follows that the Virasoro generators vanish on
the boundary. Hence, the asymptotic symmetry group of \adsd\ cannot be put
in correspondence with the conformal symmetry group of the open string when
Neumann boundary conditions are enforced. We will see in the next section that
this is due to the impossibility of realizing the symmetry in terms of local
string oscillators. Neumann boundary conditions lead to a topological theory
without local degrees of freedom and the \adsd\ asymptotic symmetry group can
be realized uniquely by the charges \cite{CM,CM3}. In this case the \adsd/CFT
correspondence is local/topological.
\section{Mode expansion and holography}
The \adsd/CFT correspondence can be realized using local oscillator degrees of
freedom as well. Let us expand the string field in normal modes
\be\lb{h1}
X^{\mu}=x^{\mu} -ip^{\mu}\log|z|^{2}+i\left(\alpha'\over 2\right)^{1/2}
\summ{1\over m}\left(\alpha^{\mu}_{m}z^{-m}+
\talpha^{\mu}_{m}\bar z^{-m}\right)\,.
\ee
Substituting the previous expansion in Eqs.\ (\ref{e17}) and comparing the
result to the expansion of the fields $M$ and $\psi$
\be\lb{h2}
M=\sumk\summ M_{k,m}x^{k}t^{m}\,,\qquad
\psi= \sum_{k=1}^{+\infty}\summ\psi_{k,m}x^{k}t^{m}\,,
\ee
we find (we assume $t>0$ for simplicity)
\ba\lb{h3}
\alpha^{\mu}_{m}&=&-i\sqrt\pi 2^{-1/2-m}\left[mM_{0,-m}-M_{1,-1-m}\pm
\psi_{1,-1-m}\right]\,,\nonumber\\
\talpha^{\mu}_{m}&=&-i\sqrt\pi 2^{-1/2-m}(-1)^m\left[mM_{0,-m}+M_{1,-1-m}\mp
\psi_{1,-1-m}\right]\,,
\ea
where the $\pm$ signs refer to the 0 and 1 components of $\alpha^{\mu}_{m}$,
respectively. Equation (\ref{h3}) puts the modes of the string in a one-to-one
correspondence with the ``gravitational'' modes of the fields $M$ and $\psi$.

Let us now enforce Dirichlet and Neumann boundary conditions on the string
field. Neumann boundary conditions imply $M_{k,m}=0$, $\psi_{k,m}=0$ for $k\ge
1$, so Eq.\ (\ref{h3}) becomes
\be\lb{h4}
\alpha^\mu_m=(-1)^{m}\talpha^\mu_m=
-i \sqrt\pi 2^{-1/2-m}\left( mM_{0,-m}\right)\,.
\ee
The generators of the Virasoro algebra of CFT$_2$ vanish identically,
\be\lb{h5}
L^{CFT}_m={1\over 2}\sumn\alpha^\mu_{m-n}\alpha_{\mu n}=0\,.
\ee
Therefore, the \adsd/CFT duality cannot be realized in terms of local
oscillators. This result has a natural interpretation. The gravitational theory
with Neumann boundary conditions is a topological theory with no local degrees
of freedom. The two-dimensional CFT action depends only on $M_{1,-m}$ and
$M_{k,m}$, $\psi_{k,m}$ with $k>1$, so vanishes at any order of the expansion.

The Dirichlet boundary conditions imply
\be\lb{h6}
p^{\mu}=0\,,\qquad \alpha^\mu_m=(-1)^{m+1}\talpha^\mu_m\qquad\to\qquad
M_{0,m}=0\quad\hbox{for}\quad m\neq 0\,,
\ee
and Eq.\ (\ref{h3}) becomes
\be\lb{h7}
\alpha^\mu_m=i\sqrt{\pi}2^{-1/2-m}
\left[M_{1,-1-m}\mp\psi_{1,-1-m}\right]\,.
\ee
The Virasoro generators of CFT$_2$ are
\be\lb{h8}
L^{CFT}_m={1\over2}\sumn\alpha^\mu_{m-n}\alpha_{\mu n}=
 -\pi 2^{-m}\sumn M_{1,-1-n}\psi_{1,-1-m+n}\,.
\ee

The previous results have two important consequences. Firstly, we see that the
mass field $M$ is constant on the boundary. This is due to the breaking of
translational invariance on the boundary that follows from the Dirichlet
conditions. [See also Eq.\ (\ref{e19}).] $M_{0,0}$ is essentially the conserved
charge and does not appear in the definition of the string modes. Secondly,
with a bit of algebra it can be proved that the gravitational modes $M_{k,m}$
and $\psi_{k,m}$ satisfy the two recurrency relations
\be\lb{h9}
M_{k+2,m-2}={m(m-1)\over (k+2)(k+1)}M_{k,m}\,,\qquad
\psi_{k+2,m-2}={m(m-1)\over (k+2)(k+1)}\psi_{k,m}\,.
\ee
Substituting Eq.\ (\ref{h6}) in the recurrency relations above, and recalling
that for the \adsd\ geometry $\psi_{0,m}=0$, the gravitational modes are
\be\lb{h10}
\begin{array}{lclcll}
k&=&\hbox{even}&\quad:\quad&M_{k,m}=0\,,\qquad&\psi_{k,m}=0\,;\\
k&=&\hbox{odd}&\quad:\quad&M_{k,m}\equiv M_{k,m}(M_{1,m},\psi_{1,m})\,,\qquad&
\psi_{k,m}\equiv\psi_{k,m}(M_{1,m},\psi_{1,m})\,.
\end{array}
\ee
The gravitational modes $M_{k,m}$ and $\psi_{k,m}$ are completely determined by
$M_{1,m}$ and $\psi_{1,m}$ which are, in turn, determined by the string modes
through Eq.\ (\ref{h7}). Therefore, the modes of the bosonic string determine
completely the sigma model, i.e., the full gravitational theory. The latter can
be expressed as a function of the gravitational modes $M_{k,m}$ and
$\psi_{k,m}$ by the perturbative expansion (\ref{e15}). Then, owing to Eq.\
(\ref{h7}) and Eq.\ (\ref{h10}), the perturbative expansion (\ref{e15}) can be
written as a function of the string modes. We conclude that gravity on \adsd\
is completely determined by the (interacting) Dirichlet open string. Vice
versa, the asymptotic two-dimensional gravitational modes near the boundary,
that describe boundary deformations, determine completely CFT$_2$. This is a
sort of holographic principle: the physics on the spacetime boundary determines
the properties of the theory in the bulk. However, it should be stressed that
we have here a somehow unusual realization of the holographic principle. 
Usually, in the context of the \ads/CFT correspondence we have a gravitational
theory defined on a $d$-dimensional bulk which is dual to a CFT theory living
on its $(d-1)$-dimensional boundary. In our case the picture is reversed: The
CFT open string lives on the two-dimensional bulk, whereas the gravitational
theory is completely defined by a boundary theory. We have already pointed out
that this property is related to the peculiar nature of gravity in two
spacetime dimensions, which is itself a conformal field theory.

The free bosonic string describes the off-shell dynamics of the classical black
hole with mass $m_{bh}$ (the fundamental state of the theory). Higher order
terms in the expansion (\ref{e15}) -- the interaction vertices of the string --
describe the off-shell corrections to the fundamental state due to effects of
the bulk. This result is a natural consequence of the topological nature of the
theory. It holds in the quantum theory as well, where the (on-shell)
fundamental state of a black hole with given classical mass $m_{bh}$ is given
by the eigenstate of the mass operator, whereas the sigma model describes the
(off-shell) black hole excitations (Quantum Birkhoff Theorem \cite{cav1}). We
recover the result that have previously anticipated: The Dirichlet bosonic
string describes the off-shell semiclassical properties of the theory whereas
the interaction terms in the perturbative expansion describe higher order
corrections to the off-shell dynamics that lead, in the quantum formalism, to
the gravitational corrections of the classical geometry.

It is worth noticing that the degrees of freedom involved in the correspondence
(\ref{h7}) are local, pure gauge, degrees of freedom. Both two-dimensional
dilaton gravity and string theory with two-dimensional target spacetime are
topological theories. So the gravitational modes $M_{1,k},\psi_{1,k}$ and the
string modes $\alpha_{k}^{\nu}$ describe pure gauge degrees of freedom. Since
we are dealing with pure gauge degrees of freedom, criticisms could be raised
about the relevance of the correspondence that we are discussing. In the next
section we will use the correspondence to calculate the statistical entropy of
two-dimensional black holes. The reader might argue that we are not counting
physical states of the two-dimensional bulk theory. However, this approach is
consistent as long as we restrict our discussion to topological theories such
as two-dimensional dilaton gravity and three-dimensional pure gravity theories
\cite{carlip1,strominger1}. Our treatment of two-dimensional dilaton gravity
and that of Refs.\ \cite{carlip1,strominger1} suggest the existence of a
nontrivial relation between local pure gauge degrees of freedom on the bulk and
topological degrees of freedom on the boundary. In the three-dimensional case
Carlip has found an explicit realization of this relation \cite{carlip1}. In
our case we have not been able to find a similar relation, yet the duality that
we have found provides a strong evidence in this direction. 

The discussion of this section has profound implications on the statistical 
interpretation of the thermodynamic quantities in Eq.\ (\ref{e7}). It is
commonly believed that the thermodynamic relations (\ref{e7}) hold within some
sort of semiclassical approximation to gravity. If the free open string really
describes semiclassical \adsd\ gravity then it should provide a statistical
description of the thermodynamic relations (\ref{e7}). This is indeed the case,
as we will see in the next section. 
\section{Two-dimensional black holes as open strings}
The \adsd/open string duality discussed in the previous sections allows to
interpreting the excitations of the gravitational theory, i.e., of the black
holes, in terms of the excitations of the string. Naively, we would be tempted
to use the equations of Sections V and VI to work out the correspondence
explicitly. For instance, we could try and use Eq.\ (\ref{h8}) to calculate the
mass of the string state which is associated with the gravitational excitations
described by the modes $M_{1,k},\psi_{1,k}$. Unfortunately, we do not know how
to relate explicitly the sigma model  modes with the physical parameters of the
black hole. So most of the equations of the previous sections cannot be
employed to describe black holes straightforwardly. However, the knowledge of
the exact form of the correspondence is not necessary. The bare fact that a
two-dimensional black hole has a dual description in terms of a two-dimensional
conformal field theory with central charge $c$ is sufficient to explain the
semiclassical behavior of the black hole.

The energy  of the CFT excitation is given by the eigenvalue, $m_{CFT}$, of
the Virasoro operator $L^{CFT}_{0}$, 
\be\lb{l1}
m_{bh}= \l m_{CFT}\,.
\ee
(With our conventions $m_{CFT}$ is dimensionless.) The energy-temperature and
entropy-mass relations of a two-dimensional CFT are \cite{ca},
\be\lb{l2}
m_{CFT}={\pi\over  12} \alpha' c T^{2}\,,\qquad
S_{CFT}=2\pi\sqrt{c \,m_{CFT}\over 6}\,,
\ee
respectively. The previous equations reproduce the functional behavior of the 
thermodynamic parameters of the black hole, Eq.\ (\ref{e7}). In order to show
that the thermodynamic behavior of the two-dimensional black hole has a
direct interpretation in terms of the microscopic dynamics of the
two-dimensional CFT, we must show that Eqs.\ (\ref{l2}) match Eqs.\ (\ref{e9})
exactly. This can be done by expressing the central charge $c$ associated with
the central extension of the Virasoro algebra generated by $L^{CFT}_{m}$ in
terms of the physical parameters of the two-dimensional black hole.

The central charge can be determined using its interpretation as a Casimir
energy. (See for instance Ref.\cite{pol}.) The transformation law of the
stress-energy tensor under the change of coordinates (\ref{e6}) is   
\be\lb{l3}
T\two_{ww}=(\p_{w}z)^{2}T\two_{zz}- {c\over 12} \{w,z\}(\p_{w}z)^{2}\,,
\ee
where $w=\tau +\sigma$ and $\{w,z\}$ is the Schwarzian derivative. The vacuum
energy is shifted by $l_0\to l_0 - a^{2}c/24$, where $l_0$ is the eigenvalue of
$ L^{CFT}_{0}$ which is associated to the vacuum. This shift corresponds to a
Casimir energy $E=-a^{2}c\l/24$.

The coordinate transformation (\ref{e6}) maps the \adsdz\ ground state
(\ref{e4}) into the \adsdp\ black hole (\ref{e5}) with mass
$m_{bh}=a^{2}\fo\l/2$. Because of the duality relation between the
gravitational theory and the Dirichlet string we can interpret the previous map
as the gravitational theory counterpart of the shift of $ L^{CFT}_{0}$ in
CFT$_2$  and equate the Casimir energy $E$ with $m_{bh}$. Actually, the
equation picks up a minus sign, 
\be\lb{l4}
E=-m_{bh},
\ee
because the coordinate transformation (\ref{e6})  maps observers. An observer
in the \adsdp\ vacuum sees the \adsdz\ vacuum filled with thermal radiation
with negative flux \cite{CM2}. Using Eq.\ (\ref{l4}) one easily finds
\be\lb{l5}
c=12\fo\,.
\ee
Inserting Eq.\ (\ref{l5}) into Eqs.\ (\ref{l2}), expressing the string length
in terms of  $\l$, $ \alpha' =2\pi/\l^{2}$, and eventually using Eq.\
(\ref{l1}), Eqs.\ (\ref{l2}) reduce to Eqs.\ (\ref{e7}).

The statistical interpretation of the two-dimensional \adsd\ black hole entropy
by means of the two-dimensional conformal theory confirms the \adsd/CFT$_{2}$
duality while stressing the peculiarity of the two-dimensional case in the
AdS/CFT family. In the case under consideration the duality maps theories that
live in spacetimes of identical dimensionality. So the relation is not
holographic in the usual sense, because it does not imply the huge reduction of
the number of degrees of freedom which is typical of the holographic principle.

This result is also understood through a different, albeit related,  argument.
The holographic principle puts an upper bound to the information that can be
encoded in a spacetime region. This upper bound is given by the 
Bekenstein-Hawking entropy
\be\lb{l6}
S_{bh}= {A\over 4 G}\,,
\ee
where $A$ is the area of the boundary surrounding the region (the area of the
black hole horizon) and $G$ is the Newton constant. In our two-dimensional case
the entropy can be written
\be
S_{bh}=2\pi \phi_{h}=2\pi r_{h}\l \phi_{0}\,,
\ee
where $\phi_{h}$ and $r_{h}$ are the dilaton and the radius evaluated at the
horizon, respectively. Since $\phi_{0}^{-1}$ plays the role of the
two-dimensional Newton constant, the previous relation can be rewritten as
$S_{bh}=2 \pi A\l/G_2$, where $A=r_{h}$. This relation is interpreted as an
information bound rather than as a holographic bound. Indeed, $A$ is not the
area of the boundary -- in our case the boundary is a point --  but the area of
the two-dimensional bulk region $0<r\le r_{h}$. Note that the previous
arguments are only valid for the Dirichlet open string. When Neumann boundary
conditions are imposed the realization of the \ads/CFT duality is more
problematic. In this case a consistent \ads/CFT duality can be realized
exclusively by a one-dimensional CFT on the boundary which supports the
conventional notion of holography. On the other hand the two-dimensional
Stefan-Boltzmann law (\ref{e7}) seems to rule out a realization of the
conformal symmetry on the boundary by means of a quantum mechanical system.

The Hawking evaporation process of the two-dimensional \adsd\ black hole
\cite{CM2} has a natural interpretation in the context of the \adsd/CFT$_{2}$
correspondence as well. In Section V we pointed out that in the $(t,x)$
coordinate frame $L^{AdS}_{0}$ generates dilatations, whereas in the coordinate
frame $(\t,\sigma)$ generates time translations. The coordinate transformation
(\ref{e6}) maps the \adsdz\  ground  state (\ref{e4}) into the \adsdp\ black
hole (\ref{e5}). Since positive frequency modes of a quantum field with respect
to Killing vector $\partial_{t}$ are not positive frequency modes with respect
to Killing vector $\partial_{\t}$, the \adsdp\ vacuum state appears filled with
thermal radiation to an observer in the \adsdz\ vacuum. The particle spectrum
can be obtained calculating the Bogoliubov  coefficients between the two vacua
\cite{CM2}. One finds that an observer in the \adsdz\ vacuum detects a thermal
flux of particles with Planck spectrum and temperature (\ref{e7}). The value of
the total Hawking flux has been calculated in Ref.\ \cite{CM2}. Therefore, the
Hawking evaporation effect emerges in the CFT context by requiring that
$L^{AdS}_{0}$ is the generator of time translations.  

Up to now we have restricted our considerations to the JT model. Our results
can be extended to the general dilaton gravity model (\ref{e1}) provided that
its solutions behave asymptotically as in Eq.\ (\ref{f1a}) and Eq.\
(\ref{f1b}). A sufficient condition is that the potential $V(\phi)$ in Eq.\
(\ref{e1}) behaves for $\phi\to \infty$ as \cite{CM3}
\be\lb{l6a}
V(\phi)=2\phi+\ord{\phi^{{-2}}}\,.
\ee
One can easily check that in this case the leading term in the weak-coupling
expansion (\ref{e14}) describes a free bosonic string. Moreover, Eqs.\
(\ref{e7}) describe the thermodynamic behavior of the corresponding black
solutions at the leading order in the large  $m_{bh}$ expansion \cite{CM3}.
Hence, the results obtained in Section VII for the JT model hold for the
general model (\ref{l6a}) at the leading order in the large $m_{bh}$ expansion.

Let us conclude this section with a remark concerning the relevance of our
results for four-dimensional black holes. The two-dimensional dilaton gravity
model (\ref{e1}) represents not only a toy model for studying gravitational
physics in a simplified context, but describes asymptotically flat
four-dimensional black holes in the near-horizon, near-extremal approximation
\cite{cadoni} as well. It can be showed that a class of black hole solutions of
the effective string theory whose near-horizon behavior is \adsd$\times S^{2}$,
and $\phi$ varies linearly, exist. Our derivation of the statistical entropy
applies straightforwardly to these solutions. 
\section{Discussion}
In the previous sections we have been able to work out in detail the 
correspondence between two-dimensional dilaton gravity on \adsd\  and open
strings.  Actually, the exact form of the correspondence is perturbative and
has only been studied at the leading order in the weak-coupling expansion
$\psi\to 0$. We have seen that in this regime two-dimensional dilaton
gravity has two degeneration limits that are described by open strings with 
Dirichlet and Neumann boundary conditions, respectively. Since the description
of this degeneracy is based on boundary conditions, it is, however, not
completely satisfactory. One would like to understand it in terms of different
regions in the parameter space of the theory.

The previous formulation of the \ads/CFT duality is very useful not only
because it makes direct contact with the original Maldacena conjecture
\cite{Wm}, but also because it can shed some light on several puzzling issues
of the AdS/CFT correspondence. The main point of this formulation is the
observation that the weak-coupling limit $\psi\to 0$ can be obtained in two
different ways. Since $\phi=\fo \l r$, the weak-coupling limit can be reached
by letting $\fo\to\infty$  at $r=\l^{-1}\f\fo^{-1}=$ constant. So we have two
weak-coupling regimes: {\it i)} $r>>1/\lambda=\sqrt{\alpha'/2\pi}$ and {\it
ii)} $r\sim1/\lambda=\sqrt{\alpha'/2\pi}$. Note that these weak-coupling limits
require that the weak-coupling expansions of the previous sections are written
in terms the variables $x/\fo$ and $(u+v)/\fo$, respectively. Since $\fo$ is
equal to $1/12$ of the central charge of the CFT, it counts the degrees of
freedom and is the two-dimensional analogue of $N$ in the Maldacena conjecture.
The limit {\it i)} corresponds to a one-dimensional CFT on the boundary and
describes the excitations of the endpoints of a Neumann open string. The limit
{\it ii)} describes a two-dimensional CFT in the bulk and describes the
excitations of a Dirichlet open string.

The duality discussed in our paper has been obtained at the zeroth order  in
the perturbative expansion (\ref{e15}). Let us discuss qualitatively how the
picture  is affected by the presence of higher order terms in Eq.\ (\ref{e15}).
Potential terms in the perturbative expansion of the sigma model do not destroy
(classical) conformal invariance. Now the model describes open strings
propagating in a curved target spacetime. The \adsd\ boundary can be regarded
as the (asymptotic) vacuum state of the theory and the string field is expanded
in normal modes around this vacuum. Imposing Dirichlet boundary conditions we
find the correspondence between the gravitational modes and the string modes,
Eq.\ (\ref{h7}). The theory can then be expressed at any order as a function of
the first order gravitational modes $M_{1,m}$, $\psi_{1,m}$ or, alternatively,
as a function of the string modes $\alpha^\mu_m$. The potential term at a given
order gives the interaction term for the modes. Obviously, dealing with an
interacting theory, the relation between the stress-energy tensor and the
Virasoro generators (\ref{f13}) is not valid. Though the first order
gravitational modes define uniquely the stress-energy tensor, the relation of
the latter with the string modes is more complicated than Eq.\ (\ref{f13}).
Consequently, we expect the central charge to be different from the central
charge of the free theory, $c=12\phi_0$, and the thermodynamic relations
(\ref{e7}) to be altered. This is no surprise. Higher order (off-shell)
corrections to the free theory on the boundary induce (quantum) corrections to
the black hole geometry that affect the derivation of the thermodynamic
relations (\ref{e7}). Calculating at a given perturbative order the black hole
geometry and the central charge one could find out how the statistical
derivation of the entropy is affected by the presence of the interaction terms.

The existence of two degeneration limits of the weak-coupled dilaton gravity
theory clarifies some controversial issues of the two-dimensional AdS/CFT
correspondence. The two-dimensional CFT with Dirichlet boundary conditions
gives a consistent explanation of the features of the dilaton gravity theory.
We might conclude that the microscopic dynamics of two-dimensional black holes
is fully captured by a two-dimensional CFT. This conclusion is not completely
satisfactory, however, because a one-dimensional CFT living on the  boundary of
\adsd\ emerges in our picture as well. The role of the AdS$_{2}$/CFT$_{1}$
correspondence, and its relation to the AdS$_{2}$/CFT$_{2}$ duality, are not
yet fully understood and deserve further investigations. In this respect, the
results of this paper seem to give contradictory indications. Although a
CFT$_{1}$ fits naturally in our scheme, it is indeed very difficult to
understand how it could explain the energy-temperature relation (\ref{e7}). 

It has been proposed \cite{g3} that conformal mechanics, possibly in the form
of large $N$ Calogero models, describes the ground state of the two-dimensional
black holes arising as near-horizon geometry of the four-dimensional
Reissner-Nordstr\"om black hole, which is characterized by a constant dilaton.
If we could extend this proposal to our case, which is characterized by a
non-constant dilaton, the conformal mechanics would describe the ground state,
whereas the two-dimensional CFT would describe the black hole excitations
\cite{g5,ms,hs}. However, the existence of a mass gap that separates the ground
state from the continuous part of the spectrum -- typical of the
Reissner-Nordstr\"om-like black holes but absent in the JT case --  seems a
crucial missing ingredient to make this proposal feasible.
\section*{Acknowledgments}
We thank D.\ Klemm, M.\ Lissia, S.\ Mignemi and P.\ Carta for useful
discussions.  M.\ Cavagli\`a is supported by a FCT grant Praxis XXI - Forma{\c
c}{\~a}o Avan{\c c}ada de Recursos Humanos, Subprograma Ci{\^e}ncia e
Tecnologia do $2^o$ Quadro Comunit{\'a}rio de Apoio, contract number
BPD/20166/99.


\begin{references}

\bibitem {Wm}
J.\ Maldacena, Adv.\ Theor.\ Math.\ Phys.\ {\bf 2} (1998) 231;
E.\ Witten, Adv.\ Theor.\ Math.\ Phys.\ {\bf 2} (1998) 253;
E.\ Witten, Adv.\ Theor.\ Math.\ Phys.\ {\bf 2} (1998) 505.

\bibitem{g1}
A.\ Strominger, \JHEP  {\bf 01} (1999) 007.

\bibitem{CM}
M.\ Cadoni and S.\ Mignemi, \PRD {\bf D59} (1999) 081501.

\bibitem{CM3}
M.\ Cadoni and S.\ Mignemi, \NPB {\bf B557} (1999) 165.

 \bibitem{CM1}
M.\ Cadoni and S.\ Mignemi, ``Symmetry Breaking, Central Charges and the
AdS$_{2}$/CFT$_{1}$ Correspondence'', [{\tt hep-th/0002256}].

\bibitem{g2}
J.\ Michelson and A.\ Strominger, \JHEP {\bf 9909} (1999) 005;
M.\ Spradlin and A.\ Strominger, \JHEP {\bf 9911} (1999) 021;
J.\ Maldacena, J.\ Michelson and A.\ Strominger, \JHEP {\bf 9902} (1999) 011;
J.\ Navarro-Salas and P.\ Navarro, \NPB {\bf B57} (2000) 250.

\bibitem{g3}
G.W.\ Gibbons and P.K.\ Townsend, \PLB {\bf B454} (1999) 187;

\bibitem{g4}
S.\ Cacciatori, D.\ Klemm and D.\ Zanon, [{\tt hep-th/9910065}].

\bibitem {g5}
S.\ Cacciatori, D.\ Klemm, W.A. Sabra and D.\ Zanon,
[{\tt hep-th/0004077}].

\bibitem{cav1} M.\ Cavagli\`a, \PRD {\bf D59 }(1999) 084011, [{\tt
hep-th/9811059}].

\bibitem{CM2}
M.\ Cadoni and S.\ Mignemi, \PRD {\bf D51 }(1995) 4319.

\bibitem{CC}
M.\ Cadoni and  M.\ Cavagli\`a, ``2D black holes as open strings: 
A new realization of the AdS/CFT duality'', [{\tt hep-th/0005179}].

\bibitem{dg}
See e.g.\ R.\ Jackiw, in: {\it Proceedings of the Second Meeting on Constrained
Dynamics and Quantum Gravity} [\NPBP 57, 162 (1997)];  M.\ Cavagli\`a, in: {\it
Proceedings of the Sixth International Symposium on Particles, Strings and
Cosmology PASCOS-98}, edited by P.\ Nath (World Scientific, Singapore, 1999)
[{\tt hep-th/9808135}]; M.\ Cavagli\`a, in: {\it Particles, Fields \&
Gravitation}, edited by J.\ Rembielinski, AIP Conf.\ Proc.\ No.\ 453 (AIP,
Woodbury, NY, 1998), pp.\ 442-448 [{\tt hep-th/9808136}].

\bibitem{JT}
C.\ Teitelboim, in: {\it Quantum theory of Gravity}, edited by S.M.\
Christensen (Hilger, Bristol); R.\ Jackiw, ibid.

\bibitem{mass}
R.B.\ Mann, \PRD {\bf D47 }(1993) 4438;
D.\ Louis-Martinez and G.\ Kunstatter,
\PRD D52 (1995) 3494;
D.\ Louis-Martinez, J.\ Gegenberg and G.\ Kunstatter,
\PLB {\bf B321 } (1994) 193;
A.T.\ Filippov, \MPL  {\bf A11 } (1996) 1691; \IJMP {\bf A12} (1997) 13.

\bibitem{MTW} See, e.g., C.W.\ Misner, K.S.\ Thorne and J.A.\ Wheeler,
{\it Gravitation} (W.H.\ Freeman and Co., New York, 1973). 

\bibitem{RT}
T.\ Regge and C.\ Teitelboim, \ANP 174 (1974) 463;
A.\ Hanson, T.\ Regge and C.\ Teitelboim, {\it Constrained Hamiltonian
Systems} (Accademia Nazionale dei Lincei, Roma, 1976).

\bibitem {pol} J.\ Polchinski, {\it String Theory} (Cambridge Univ.\ Press,
Cambridge UK, 1998).

\bibitem{cav2} M.\ Cavagli\`a and C.\ Ungarelli, \PRD {\bf D61} (2000) 064019, 
[{\tt hep-th/9912024}].

\bibitem{Cadoni}
M.\ Cadoni, \PLB {\bf B395} (1997) 10.

\bibitem{carlip1}
S.\ Carlip, \PRD {\bf D51} (1995) 632.

\bibitem{strominger1}
A.\ Strominger, \JHEP {\bf 02 } (1998) 009. 

\bibitem{BD}
N.D.\ Birrel and P.C. Davies, {\it Quantum Fields in Curved Spaces} (Cambridge
Univ.\ Press, Cambridge UK, 1982).

\bibitem {ca}
J.A.\ Cardy, \NPB {\bf B270} (1986) 186.

\bibitem {cadoni}
M.\ Cadoni, \PRD  {\bf D60} (1999) 084016.

\bibitem {ms}
J.J.\ Maldacena and  A.\ Strominger, \PRD {\bf D56} (1997) 4975. 

\bibitem {hs}
G.T.\ Horowitz and A.\ Strominger, \PRL {\bf 77} (1996) 2368.

\end{references}
\end{document}